\documentclass[11pt,twoside]{article}
\usepackage{asp2014}

\hypersetup{colorlinks=false}

\aspSuppressVolSlug
\resetcounters

\begin{document}

\title{Constraining the Initial-Final Mass Relation with Wide Double White Dwarfs}
\author{Jeff J.\ Andrews,$^1$ Marcel A.\ Ag\"ueros,$^1$ Alexandros Gianninas,$^2$ Mukremin Kilic,$^2$ Saurav Dhital,$^3$ and Scott Anderson$^4$
\affil{$^1$Columbia University, New York, NY, USA; \email{andrews@astro.columbia.edu}}
\affil{$^2$University of Oklahoma, Norman, OK, USA}
\affil{$^3$Embry-Riddle Aeronautical University, Daytona Beach, FL, USA}
\affil{$^4$University of Washington, Seattle, WA, USA}}

% This section is for ADS Processing.  There must be one line per author.
\paperauthor{Jeff J.\ Andrews}{andrews@astro.columbia.edu}{}{Columbia University}{Department of Astronomy}{New York}{NY}{10027}{USA}
\paperauthor{Marcel A.\ Ag\"ueros}{marcel@astro.columbia.edu}{}{Columbia University}{Department of Astronomy}{New York}{NY}{10027}{USA}
\paperauthor{Alexandros Gianninas}{alexg@nhn.ou.edu}{}{University of Oklahoma}{Department of Physics and Astronomy}{Norman}{OK}{73019}{USA}
\paperauthor{Mukremin Kilic}{kilic@ou.edu}{}{University of Oklahoma}{Department of Physics and Astronomy}{Norman}{OK}{73019}{USA}
\paperauthor{Saurav Dhital}{Saurav.Dhital@erau.edu}{}{Embry-Riddle Aeronautical University}{Department of Physical Sciences}{Daytona Beach}{FL}{32114}{USA}
\paperauthor{Scott Anderson}{sfander@u.washington.edu}{}{University of Washington}{Department of Astronomy}{Seattle}{WA}{98195}{USA}

\begin{abstract}
In wide double white dwarf (DWD) binaries, in which the co-eval WDs evolve independently, the more massive, faster-evolving WD can be used to obtain a main-sequence lifetime for the less-massive WD. By converting this lifetime into an initial mass for the less-massive WD, and combining it with the spectroscopically derived mass for this WD, one can constrain the initial-final mass relation (IFMR). However, the number of known and well-characterized DWDs is small, severely limiting their usefulness for this analysis. To obtain new constraints on the IFMR, we search for wide DWDs in the Sloan Digital Sky Survey (SDSS) Data Release 9. We find 65 new candidate systems, thereby raising the number of known wide DWDs to 142. We then engage in a spectroscopic campaign to characterize these pairs, identifying 32 DA/DA pairs, two DA/DB pairs, four DA/DAH candidate pairs, a previously unidentified candidate triple WD system, and five DA/DC WDs. We present a reanalysis of the constraint on the IFMR placed by \citet{finley97} using the DWD PG 0922$+$162, and finish by discussing how it could be expanded to a generic set of wide DWDs. 
\end{abstract}

\section{Introduction}
\citet{finley97} constrained the IFMR using the wide DWD PG 0922$+$162. These authors compared the more massive WD ({\lower0.8ex\hbox{$\buildrel >\over\sim$}}1.10 M$_{\odot}$) to similarly massive WDs in open clusters for which an initial (ZAMS) mass $M_{\rm i}$ had been published, thereby obtaining $M_{\rm i} = 6.5 \pm 1.0$~M$_{\odot}$\ for this WD. \citet{finley97} converted this mass into a pre-WD lifetime of 42$-$86 Myr, to which they added the cooling age $\tau_{\rm cool}$\ of the massive WD to derive a system age of $320 \pm 32$ Myr. These authors then used the less massive WD in PG 0922$+$162 to constrain the IFMR: they derived a pre-WD lifetime for this 0.79 M$_{\odot}$\ WD of $231 \pm 34$ Myr by subtracting its $\tau_{\rm cool}$\ from the system age, and obtained $M_{\rm i} = 3.8 \pm 0.2$~M$_{\odot}$\ for its progenitor. The \citet{finley97} result is one of the most stringent constraints on the IFMR, and is one of the reasons \citet{weidemann00} anchored his semi-empirical IFMR at $M_{\rm i} =4.0$~M$_{\odot}$\ and $M_{\rm WD} = 0.80$~M$_{\odot}$.

In principle, if a reasonable estimate of the system age can be determined from the more massive WD, this method can be applied to \emph{any} wide DWD. The \citet{finley97} study has not been widely replicated, however. Until recently, there were $\approx$35 known wide DWDs; many of these lacked spectra, and even those with spectra were ill-suited to this analysis because of large $\tau_{\rm cool}$\ uncertainties. In \citet[][]{andrews12} we identified 11 new DWDs selected by searching for common proper motion ($\mu$) companions to spectroscopically confirmed SDSS WDs. This pilot study illustrated the robustness of our method for selecting DWDs. Our tests to separate random alignments from high-confidence pairs indicated that wide DWDs could be identified without spectroscopy, using SDSS photometry and a minimum $\mu$ cut instead. Here, we present the findings of our expanded search for DWDs in the SDSS Data Release 9. We identified 65 systems, bringing the total number of candidate wide DWDs to 142. We then discuss how spectroscopic follow-up may allow us to use these DWDs to constrain the IFMR. 

\section{Searching for Wide DWDs}\label{search}
\subsection{Common Proper Motion Pairs} \label{search:dr9}
We first searched for DWDs by matching proper motions of WD candidates in the SDSS DR9. From DR9's $>$$9\times10^8$ primary photometric objects, we selected those classified as stars that matched our photometric and $\mu$ quality constraints ($\sigma_{ugr} < 0.15$ mag, $\sigma_{iz} < 1.0$ mag, $\mu > 35$ mas~yr$^{-1}$, $\sigma_{\mu} < 10$ mas yr$^{-1}$). To include both hydrogen-atmosphere DA WDs and helium-atmosphere DB WDs in this search, we used a liberal color constraint, selecting those stars with $-0.7 < (g-r) < 0.4$ and $-0.7 < (u-g) < 0.75$. The left panel of Figure \ref{fig:DR9_search} shows the $(u-g)$ versus $(g-r)$ colors of the $\approx$$4\times10^4$ SDSS objects that met our quality constraints and fell within this region of color space.

Our $\mu > 35$~mas~yr$^{-1}$ criterion should eliminate nearly all QSOs. Contaminating main-sequence stars and subdwarfs are more difficult to remove, as these objects may overlap with WDs in color and may have $\mu > 35$ mas~yr$^{-1}$. However, WDs can be effectively separated from blue stars in a reduced proper motion (H$_{\rm r}$) diagram. The right panel in Figure~\ref{fig:DR9_search} is H$_{\rm r}$ versus $(g-i)$ for the objects in our sample. We used the dashed line in the right panel of Figure~\ref{fig:DR9_search}, adapted from \citet{smith09}, to separate subdwarfs from WDs, thereby reducing our sample to $\approx$34,000 objects.

Next, we searched for $\mu$ matches. A match occurs when two WDs had an angular separation $\theta < 5^{\prime}$, and, following \citet{dhital10}, a matching parameter $\Sigma^2 < 2$.\footnote{$\Sigma^2 =\left( \Delta \mu_{\alpha}/\sigma_{\Delta \mu_{\alpha}} \right)^2 + \left( \Delta \mu_{\delta}/\sigma_{\Delta \mu_{\delta}} \right)^2 $, where $\Delta \mu$ is the difference in $\mu$ in right ascension ($\alpha$) and declination ($\delta$), and $\sigma_{\mu}$ is the error in $\mu$.} We identified 57 candidate DWDs, and selected the 36 pairs with $\theta < 100^{\prime\prime}$\ as high-probability DWD candidates; 13 had previously been identified. We added four pairs at larger $\theta$ but with $\mu>80$ mas yr$^{-1}$; with such large common $\mu$ these are unlikely to be random alignments. The number of new candidate common $\mu$ DWDs is 27.  

\articlefiguretwo{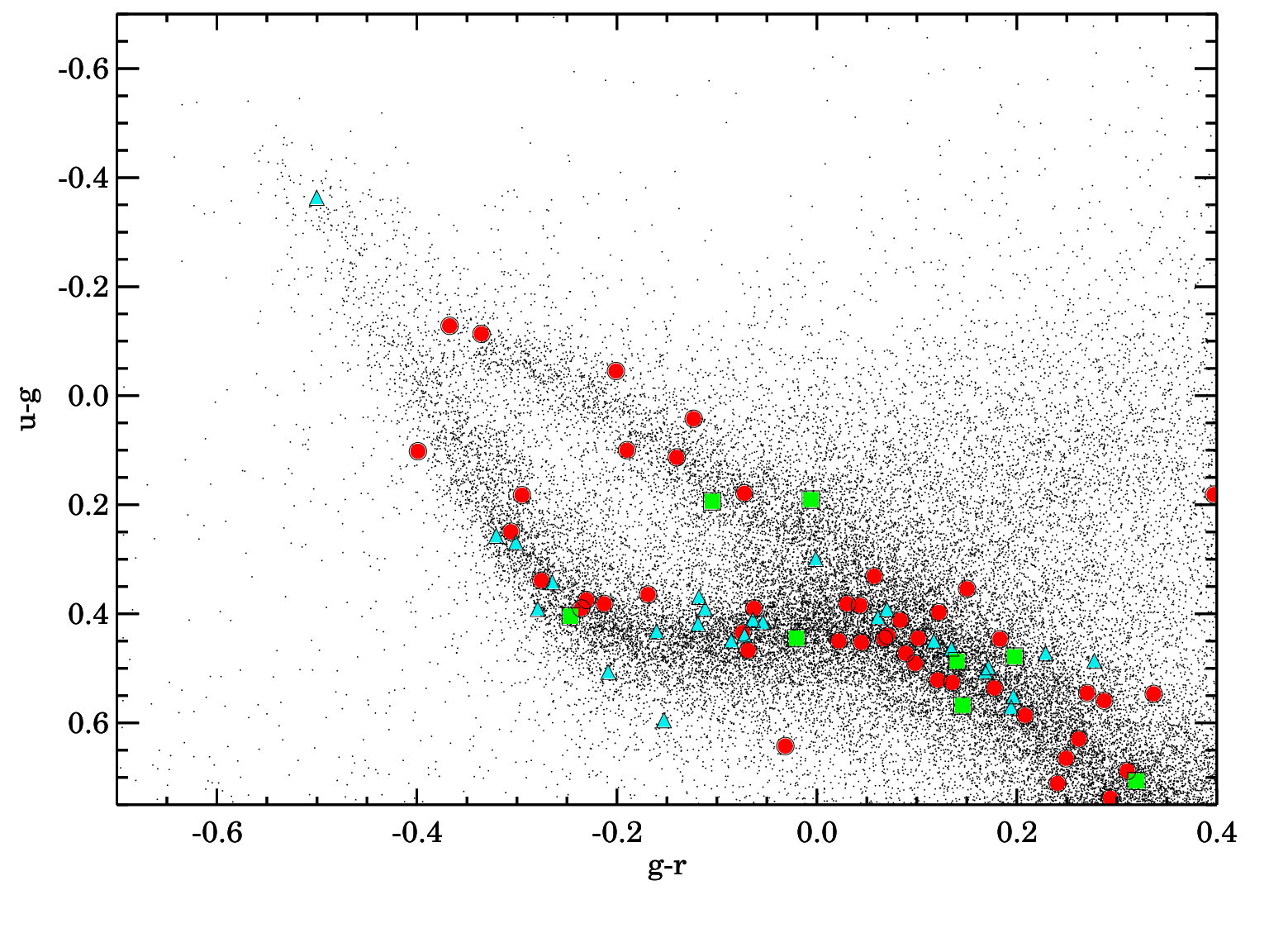}{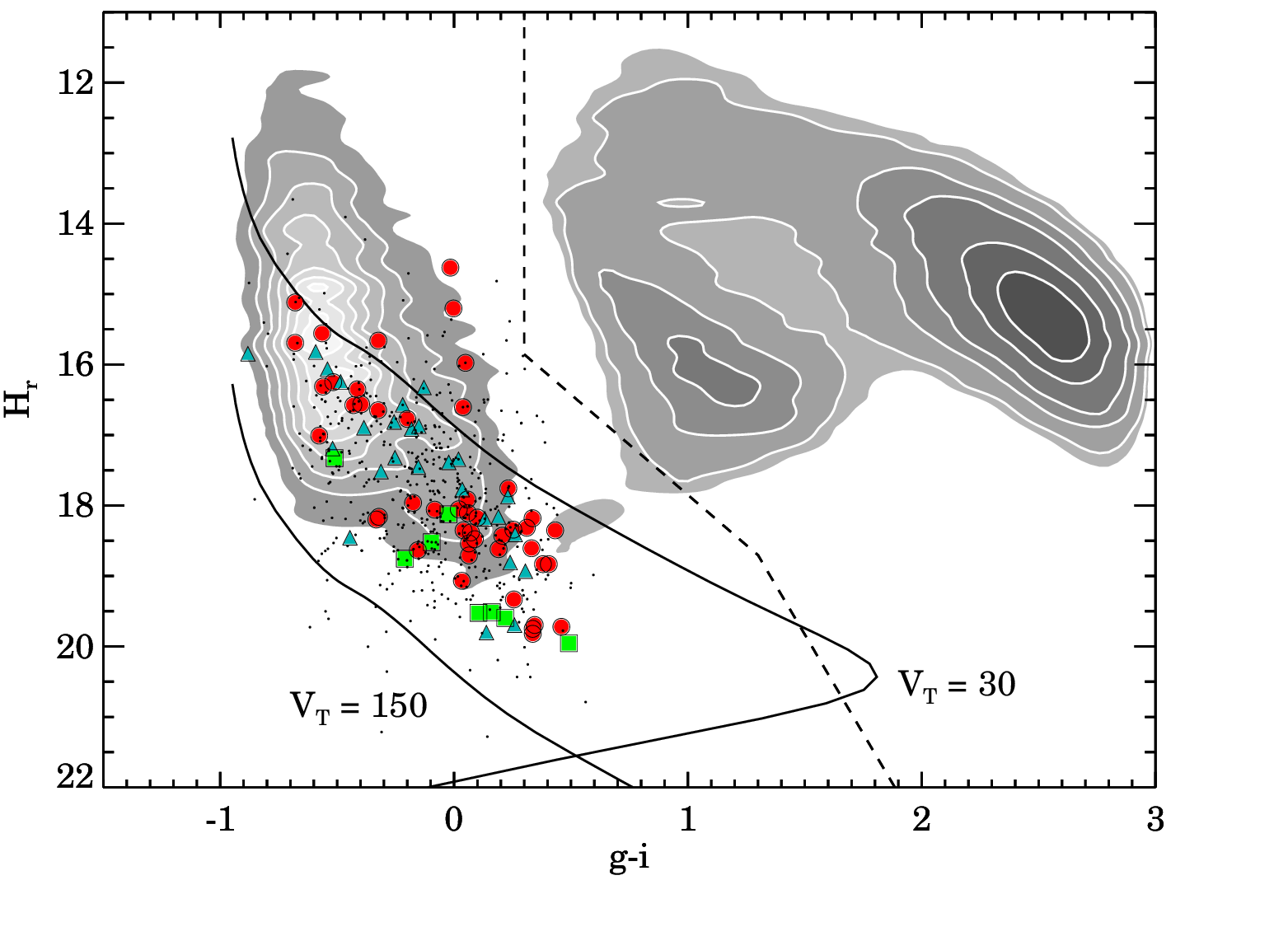}{fig:DR9_search}{The left panel shows the $(u-g)$ versus $(g-r)$ for $\approx$$4\times10^4$ SDSS objects that pass our photometric and $\mu$ quality constraints. The sickle-shaped band corresponds to the DA cooling sequence while DB WDs lie in a parallel band to the upper right of this band. The right panel shows reduced proper motion (H$_{\rm r}$) versus $(g-i)$ for the same objects. The loci at $(g-i) \approx -0.5$, 1, and 2.5 indicate the population of WDs (from the \citet{kleinman13} SDSS catalog), subdwarfs and main-sequence stars, respectively. We used the dashed line, adapted from \citet{smith09}, to separate subdwarfs from candidate WDs. The solid lines represent the locations of WDs for transverse velocities V$_{\rm T}$ = 30 km~s$^{-1}$ (corresponding to the disk population) and 150 km~s$^{-1}$ (the halo population). The symbols are the same in both panels: the circles are the candidate WDs in 23 new candidate wide DWDs identified here, the squares are candidate WDs in four additional new candidate systems, and the triangles are the WDs in the 13 known systems we re-detect. }

\subsection{Astrometrically Close Pairs} \label{search:no_pm}
Dhital et al.\ (in prep.) search SDSS for photometrically resolved, small $\theta$ pairs of low-mass stars. These authors identify $>$4$\times10^4$ pairs with $\theta = 0.4$$-$10$^{\prime\prime}$\ and argue that wide pairs can be efficiently identified without having to match $\mu$. Similarly, \citet{baxter14} identified a set of wide SDSS DWDs with $\theta\ {\lower0.8ex\hbox{$\buildrel <\over\sim$}}\ 30^{\prime\prime}$\ based only on photometry.

We began our photometric search for DWDs by applying the more stringent color-color constraint described in \citet[][]{girven11} to obtain candidate WDs: without $\mu$ measurements, subdwarfs and QSOs become a significant source of contamination. \citet{girven11} estimated that 17\% of the objects falling within this color-color region are QSOs. However, this was based on a $g<19$ mag sample of objects with SDSS spectra. Our sample extends to $g=21$, and we expected QSOs to be a more significant contaminant. We therefore added the additional constraint that objects with $g>18$ should have $(u-g)<-0.1$. Of the surviving objects, we selected the 43 pairs with $\theta < 10^{\prime\prime}$\ as high-confidence candidate DWDs. Five of these were previously known, so that we have 38 new, photometrically selected, candidate DWDs.

Among the previously known wide DWDs are the 11 systems identified in \cite{andrews12} and 36 systems identified elsewhere in the literature. To that sample we add 27 common $\mu$ systems and 38 identified by their small $\theta$. \citet{baxter14} found 53 wide DWDs in SDSS and spectroscopically confirmed 26 of these (one additional pair was found to be a contaminant). Thirty of the 53 DWDs are new, and 11 of these are spectroscopically confirmed. In total, we therefore have a sample of 142 candidate and confirmed wide DWDs, including 19 \citet{baxter14} new candidate pairs and the 11 these authors have confirmed spectroscopically.

\section{Spectroscopically Characterizing DWDs}\label{spec}
Over 13 half nights between 2012 Sep and 2013 Sep, we observed 34 DWDs with the
3.5-m telescope at Apache Point Observatory, NM\footnote{The Apache Point Observatory $3.5$-m telescope is operated by the Astrophysical Research Consortium.} with the Dual Imaging Spectrograph in its high-resolution mode (R $\approx 2500$ at H$\beta$). To these spectra we add $\approx$30 spectra from SDSS (R $\approx$ 1800) and $\approx$10 spectra taken with Very Large Telescope (R $\approx$ 15,000) from the Supernova Progenitor Survey (D.~Koester, pers.~comm.). In total, we have 114 spectra for 97 WDs in wide DWDs. Our contamination rate by non-WDs is extremely low ({\lower0.8ex\hbox{$\buildrel <\over\sim$}}5\%): one target is an A star.

To obtain $T_{\rm eff}$\ and log~$g$\ for these WDs, we used the technique developed by \citet{bergeron92} and described in \citet[][and references therein]{gianninas11}, which incorporates model atmospheres for WDs with $6.5 \leq$ log~$g$\ $\leq 9.5$. The observed and theoretical spectra are normalized to a continuum set to unity, and the observed  H$\beta$ to H8 lines are fit simultaneously to the synthetic spectra. We applied the fitting formulas \citet{tremblay13} provide to the $T_{\rm eff}$\ and log~$g$\ solutions to adjust from the standard mixing-length approximation to the 3D simulation results. Next, we used the \citet{tremblay11} models to map our $T_{\rm eff}$\ and log~$g$\ values to $\tau_{\rm cool}$\ and masses ($M_{\rm WD}$) for each of our WDs. Our fits also provided distances to the WDs, determined by comparing photometric magnitudes with absolute magnitudes from the spectroscopic solutions.

\section{Adapting and Applying the Finley \& Koester (1997) Method}\label{meth}
In Figure~\ref{fig:PG0922} we reproduce one of the critical steps used by \citet{finley97} to constrain the IFMR: the conversion of stellar main-sequence lifetimes into initial masses. Because the relation between lifetime and mass is steeper for longer-lived/lower-mass main-sequence stars, even a large uncertainty in the assumed $M_{\rm i}$ for the more massive WD in PG 0922$+$162 results in a stringent constraint on the $M_{\rm i}$ of the less massive WD.

To adapt this method to a generic wide DWD, we must be able to estimate the $M_{\rm i}$ of the more massive WD in the pair. Instead of comparing to WDs in clusters, as was done by \citet{finley97}, we can use the approximate IFMRs provided by stellar evolution codes. Such theoretical relations are imperfect, primarily due to the difficulty of modeling convective overshooting and asymptotic giant branch winds. Nevertheless, by varying the different prescriptions for these processes and comparing to simulations from different groups, we can estimate the uncertainty in these theoretical relations. For the majority of the WD mass range, these uncertainties are large, leading to a wide range for $M_{\rm i}$ for the more massive WD. However, Figure~\ref{fig:PG0922} shows that even large uncertainties in this estimate can lead to important constraints on the IFMR. We are in the process of computing sets of theoretical relations with {\tt MESA}.

\articlefigure{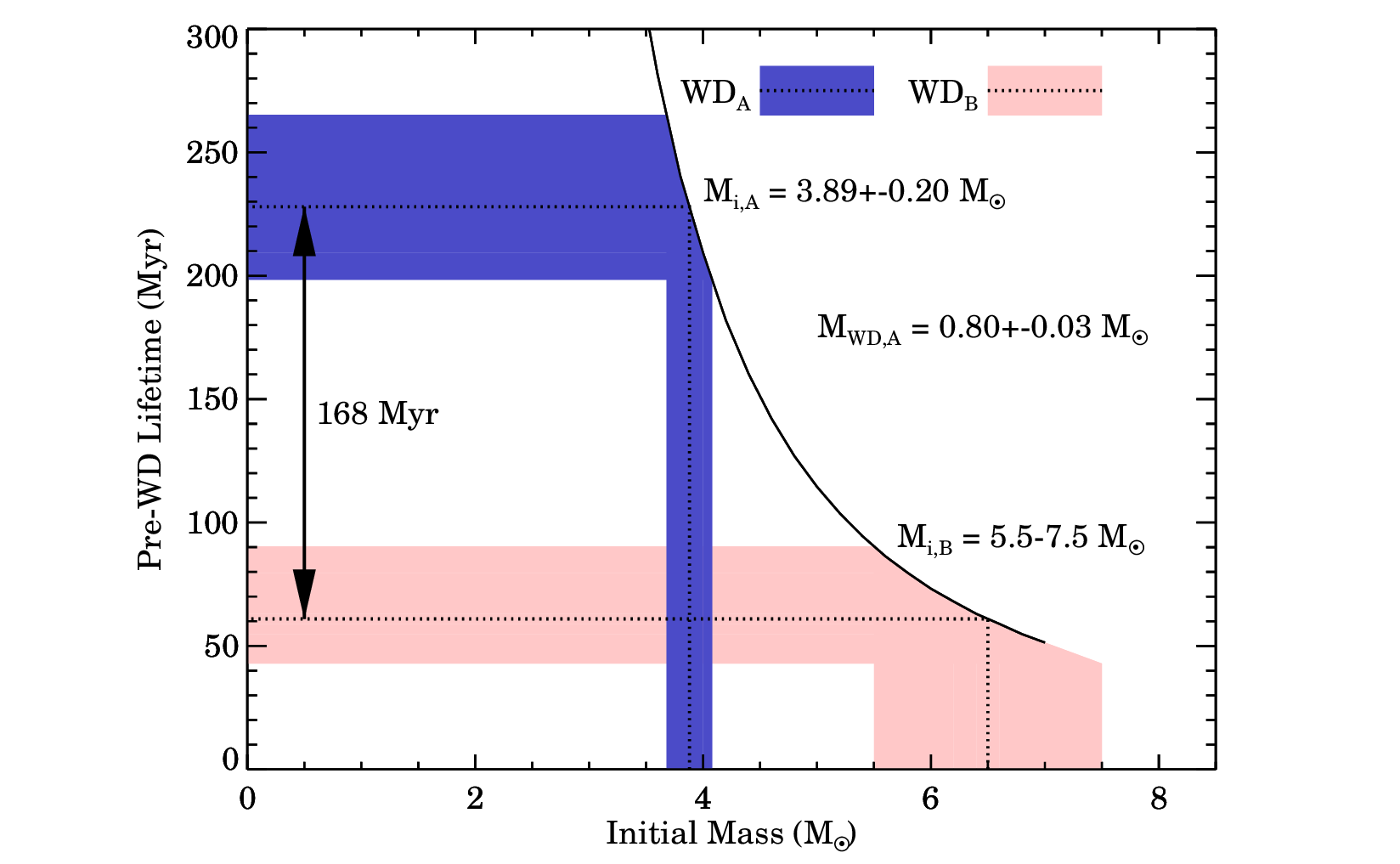}{fig:PG0922}{Converting initial masses into pre-WD lifetimes. \citet{finley97} assigned the more massive WD$_{\rm B}$ in PG~0922$+$162 a $M_{\rm i} = 6.5 \pm 1.0$~M$_{\odot}$, which we translate into a pre-WD lifetime of 43$-$90 Myr with {\tt MESA}. Adding the $\tau_{\rm cool}$\ difference between the two WDs gives a pre-WD lifetime of 198$-$265 Myr for the less massive WD$_{\rm A}$. We derive a corresponding $M_{\rm i} = 3.89 \pm 0.20$~M$_{\odot}$\ for WD$_{\rm A}$; \citet{finley97} found $3.8 \pm 0.2$ M$_{\odot}$. Despite the relatively large uncertainty in the initial mass of WD$_{\rm B}$, the uncertainty in the initial mass of WD$_{\rm A}$ is small, particularly compared to typical uncertainties derived from other observational methods.}

\section{Conclusions}
We have searched SDSS DR9 for wide DWDs, identifying 65 new candidate pairs. Through our spectroscopic follow-up campaign, we characterized 32 DA/DA pairs. We repeat the analysis of \citet{finley97}, who constrained the IFMR using the wide DWD PG 0922$+$162, obtaining nearly identical results. This method can be expanded to other wide DWDs in our catalog. With the future {\it Gaia} data release, many more of these systems will likely be identified, therefore refining this method will be important in placing future constraints on the IFMR.

\acknowledgements We thank P.~Bergeron for fitting the DB WDs in our sample and D.~Koester for providing us with VLT spectra of several of the WDs discussed here. We thank Lars Bildsten, Falk Herwig, Silvia Catalan, Marcelo Miller Bertolami, and Rodolfo Montez, Jr.~for stimulating and helpful discussions. MK and AG gratefully acknowledge the support of the NSF and NASA under grants AST-1312678 and NNX14AF65G, respectively.


\begin{thebibliography}{}
\expandafter\ifx\csname natexlab\endcsname\relax\def\natexlab#1{#1}\fi
\expandafter\ifx\csname url\endcsname\relax
  \def\url#1{\texttt{#1}}\fi
\expandafter\ifx\csname urlprefix\endcsname\relax\def\urlprefix{URL }\fi
\providecommand{\eprint}[2][]{\url{#2}}


\bibitem[{{Andrews} et~al.(2012){Andrews}, {Ag{\"u}eros}, {Belczynski},
  {Dhital}, {Kleinman}, \& {West}}]{andrews12}
{Andrews}, J.~J., {Ag{\"u}eros}, M.~A., {Belczynski}, K., {Dhital}, S.,
  {Kleinman}, S.~J., \& {West}, A.~A. 2012, \apj, 757, 170. \eprint{1209.0175}

\bibitem[{{Baxter} et~al.(2014){Baxter}, {Dobbie}, {Parker}, {Casewell},
  {Lodieu}, {Burleigh}, {Lawrie}, {K{\"u}lebi}, {Koester}, \&
  {Holland}}]{baxter14}
{Baxter}, R.~B., {Dobbie}, P.~D., {Parker}, Q.~A., {Casewell}, S.~L., {Lodieu},
  N., {Burleigh}, M.~R., {Lawrie}, K.~A., {K{\"u}lebi}, B., {Koester}, D., \&
  {Holland}, B.~R. 2014, \mnras, 440, 3184. \eprint{1403.4046}

\bibitem[{{Bergeron} et~al.(1992){Bergeron}, {Saffer}, \&
  {Liebert}}]{bergeron92}
{Bergeron}, P., {Saffer}, R.~A., \& {Liebert}, J. 1992, \apj, 394, 228

\bibitem[{{Dhital} et~al.(2010){Dhital}, {West}, {Stassun}, \&
  {Bochanski}}]{dhital10}
{Dhital}, S., {West}, A.~A., {Stassun}, K.~G., \& {Bochanski}, J.~J. 2010, \aj,
  139, 2566. \eprint{1004.2755}

\bibitem[{{Finley} \& {Koester}(1997)}]{finley97}
{Finley}, D.~S., \& {Koester}, D. 1997, \apjl, 489, L79

\bibitem[{{Gianninas} et~al.(2011){Gianninas}, {Bergeron}, \&
  {Ruiz}}]{gianninas11}
{Gianninas}, A., {Bergeron}, P., \& {Ruiz}, M.~T. 2011, \apj, 743, 138.
  \eprint{1109.3171}

\bibitem[{{Girven} et~al.(2011){Girven}, {G{\"a}nsicke}, {Steeghs}, \&
  {Koester}}]{girven11}
{Girven}, J., {G{\"a}nsicke}, B.~T., {Steeghs}, D., \& {Koester}, D. 2011,
  \mnras, 417, 1210. \eprint{1106.5886}

\bibitem[{{Kleinman} et~al.(2013){Kleinman}, {Kepler}, {Koester}, {Pelisoli},
  {Pe{\c c}anha}, {Nitta}, {Costa}, {Krzesinski}, {Dufour}, {Lachapelle},
  {Bergeron}, {Yip}, {Harris}, {Eisenstein}, {Althaus}, \&
  {C{\'o}rsico}}]{kleinman13}
{Kleinman}, S.~J., {Kepler}, S.~O., {Koester}, D., {Pelisoli}, I., {Pe{\c
  c}anha}, V., {Nitta}, A., {Costa}, J.~E.~S., {Krzesinski}, J., {Dufour}, P.,
  {Lachapelle}, F.-R., {Bergeron}, P., {Yip}, C.-W., {Harris}, H.~C.,
  {Eisenstein}, D.~J., {Althaus}, L., \& {C{\'o}rsico}, A. 2013, \apjs, 204, 5.
  \eprint{1212.1222}

\bibitem[{{Smith} et~al.(2009){Smith}, {Evans}, {Belokurov}, {Hewett},
  {Bramich}, {Gilmore}, {Irwin}, {Vidrih}, \& {Zucker}}]{smith09}
{Smith}, M.~C., {Evans}, N.~W., {Belokurov}, V., {Hewett}, P.~C., {Bramich},
  D.~M., {Gilmore}, G., {Irwin}, M.~J., {Vidrih}, S., \& {Zucker}, D.~B. 2009,
  \mnras, 399, 1223. \eprint{0904.1012}

\bibitem[{{Tremblay} et~al.(2011){Tremblay}, {Bergeron}, \&
  {Gianninas}}]{tremblay11}
{Tremblay}, P.-E., {Bergeron}, P., \& {Gianninas}, A. 2011, \apj, 730, 128.
  \eprint{1102.0056}

\bibitem[{{Tremblay} et~al.(2013){Tremblay}, {Ludwig}, {Steffen}, \&
  {Freytag}}]{tremblay13}
{Tremblay}, P.-E., {Ludwig}, H.-G., {Steffen}, M., \& {Freytag}, B. 2013, \aap,
  559, A104. \eprint{1309.0886}

\bibitem[{{Weidemann}(2000)}]{weidemann00}
{Weidemann}, V. 2000, \aap, 363, 647


\end{thebibliography}
\end{document}